\documentclass[aps,prl,preprint,superscriptaddress,showpacs]{revtex4}
\usepackage{graphicx}
\usepackage{bm}

\begin{document}

\title{Chiral charge-density-waves}


\author{J. Ishioka}
\affiliation{Department of Applied Physics, Hokkaido University, Sapporo 060-8628, Japan}
\author{Y. H. Liu}
\affiliation{Department of Physics, Hokkaido University, Sapporo 060-0810, Japan}
\author{K. Shimatake}
\affiliation{Department of Applied Physics, Hokkaido University, Sapporo 060-8628, Japan}
\author{T. Kurosawa}
\affiliation{Department of Physics, Hokkaido University, Sapporo 060-0810, Japan}
\author{K. Ichimura}
\affiliation{Department of Applied Physics, Hokkaido University, Sapporo 060-8628, Japan}
\affiliation{Center of Education and Research for Topological Science and Technology, Hokkaido University, Sapporo 060-8628, Japan}
\author{Y. Toda}
\affiliation{Department of Applied Physics, Hokkaido University, Sapporo 060-8628, Japan}
\affiliation{Center of Education and Research for Topological Science and Technology, Hokkaido University, Sapporo 060-8628, Japan}
\author{M. Oda}
\affiliation{Department of Physics, Hokkaido University, Sapporo 060-0810, Japan}
\affiliation{Center of Education and Research for Topological Science and Technology, Hokkaido University, Sapporo 060-8628, Japan}
\author{S. Tanda}
\altaffiliation[Corresponding author: ]{tanda@eng.hokudai.ac.jp}
\affiliation{Department of Applied Physics, Hokkaido University, Sapporo 060-8628, Japan}
\affiliation{Center of Education and Research for Topological Science and Technology, Hokkaido University, Sapporo 060-8628, Japan}
\date{\today}

\begin{abstract}
We discovered the chirality of charge density waves (CDW) in 1T-TiSe$_2$ by using scanning tunnelling microscopy (STM) and optical ellipsometry.  We found that the CDW intensity becomes $I{a_1}:I{a_2}:I{a_3} = 1:0.7 \pm 0.1:0.5 \pm 0.1$, where $Ia_i$ (i =1, 2, 3) is the amplitude of the tunnelling current contributed by the CDWs. There were two states, in which the three intensity peaks of the CDW decrease \textit{clockwise} and \textit{anticlockwise} when we index each nesting vector in order of intensity in the Fourier transformation of the STM images. The chirality in CDW results in the three-fold symmetry breaking. Macroscopically, two-fold symmetry was indeed observed in optical measurement. We propose the new generalized CDW chirality $H_{CDW} \equiv \mbox{\boldmath $q_1$} \cdot (\mbox{\boldmath $q_2$}\times \mbox{\boldmath $q_3$})$, where $\mbox{\boldmath $q_i$}$ are the nesting vectors, which is independent of the symmetry of components. The nonzero $H_{CDW}$ - the triple-$\mbox{\boldmath $q$}$ vectors do not exist in an identical plane in the reciprocal space - should induce a real-space chirality in CDW system. 
\end{abstract}

\maketitle


Chirality is an important concept in particle physics and the condensed matter field in relation to baryogenesis \cite{dolgov} or spontaneous symmetry breaking. In condensed matter physics, many researchers have attempted to study the chiral properties of ${}^3$He-A \cite{volo} and chiral {\it p}-wave superconductors \cite{maeno,nobukane}. Is it possible to break the chiral symmetry in electron systems that also have macroscopic quantum order such as charge density waves (CDWs)?

Chirality is often characterized by using the parameter of helicity $h=\sigma \cdot \mbox{\boldmath $p$}/|p| $ (${\bf \sigma}$ :spin \mbox{\boldmath $p$} :momentum) in a photon, or $h= \mbox{\boldmath $v$} \cdot {\bf \omega}$ (\mbox{\boldmath $v$} :velocity, ${\bf \omega}$ :vorticity) in fluids. If the helicity is nonzero in a system, the system has chirality. Chirality describes the character of an object or structure that cannot be superimposed on its mirror image. So far, there is no evidence of the chirality in one-dimensional (1D) CDWs and two-dimensional (2D) CDWs, as usual, since a CDW has the form $\rho = $cos$(kx+\phi)$ ($\rho$: charge density, $k$: wavenumber, $x$: position, and $\phi$: phase), which is invariant by space inversion, or parity transformation ($k \to (-k)$ and $x \to (-x)$).

In this paper we report the chirality of a CDW in 1T-TiSe$_2$ \cite{STM1,STM2,seityou,DiSalvo,band,diffraction,excitonic,Jahn} measured directly by using STM and optical ellipsometry. STM is used to investigate the microscopic electron distribution and the phase of the CDW order parameter in real space \cite{STM1,STM2,STM_fft}. In the measurement, the CDW intensity becomes $I{a_1}:I{a_2}:I{a_3} = 1:0.7 \pm 0.1:0.5 \pm 0.1$, where $Ia_i$ (i =1, 2, 3) is the amplitude of the tunnelling current contributed by the CDWs. We then found two states with different helicities. When we index each nesting vector in order of intensity in the Fourier transformation of the STM images, the three intensity peaks of the CDW decrease \textit{clockwise} and \textit{anticlockwise}.
Moreover, the two-color pump-probe pulse method was performed to determine the macroscopic property of the relaxation of excited electrons by pump laser pulse. The direction dependence in transient reflectivity has two-fold symmetry.
Since it is impossible to explain the electron distribution using conventional CDW theory as with 1T-TaS$_2$, we propose a new CDWs configuration based on the three-dimensional (3D) nesting vectors of 1T-TiSe$_2$ in reciprocal space. In real space, a phase shift along the nesting directions can reduce the Coulomb interaction and the phase shift between three nesting vectors induces helical stacking in CDW unit cell. As a result, the CDW system has chirality based on the direction of helical stacking.

We provide the first evidence of chirality in a CDW. Moreover, we propose generalized CDW chirality in a triple-\mbox{\boldmath $q$} system, 
 
1T-TiSe$_2$ crystallizes in a 1T-CdI$_2$ type structure. The crystal is constructed of Se-Ti-Se layers. These Se-Ti-Se layers consist of a plane occupied by Ti atoms that is sandwiched between similar planes of Se atoms on either side. All the Se-Ti-Se layers are weakly accumulated by the van der Waals attraction along the {\it c}-axis. 1T-TiSe$_2$ undergoes a CDW transition into a 2{\it a$_0$}$\times2${\it a$_0$}$\times$2{\it c$_0$} super-lattice ({\it a$_0$} = 3.54 \AA, {\it c$_0$} = 6.00 \AA{} are lattice constants) at {\it T$_c$} = 200 K \cite{diffraction}.

We synthesized a single crystal of 1T-TiSe$_2$ using the self-vapor transportation method \cite{seityou}. By controlling the growth temperature, we obtained sample A at 700 $^\circ\mathrm{C}$ and sample B at 800 $^\circ\mathrm{C}$. To check the CDW phase transition, we measured the temperature dependence of the resistivity of each sample. The resistivity behavior of each sample was consistent with that described in previous reports \cite{DiSalvo}. 

In the STM measurements, the samples were cleaved {\it in situ} just before the approach of the STM tip toward the surface in an ultra-high vacuum at 77 K. We were able to observe atomically resolved STM images in the constant height mode with a constant sample voltage $V_s$ applied between the tip and the sample. The STM images of sample A and B were obtained at 84 and 6.3 K, respectively.

The two color pump-probe pulse method was employed with the cleaved TiSe$_2$ surface using a micro optical setup with a spacial resolution of approximately 10 $\mu m$ \cite{shimatake}. We used an optical parametric oscillator pumped by a mode locked Ti:sapphire laser (76 MHz repetition rate). The time resolution was estimated to be 200 fs. The probe energy was variable in the 1.00 to 1.77 eV range and sufficient to excite an electron beyond the CDW gap. The pump and probe pulse polarization was varied to investigate the dominant direction of the deviation of reflectivity $\Delta R(t)$. The sample was mounted on a helium-flow cryostat and the reflection was measured at 3K.

\begin{figure}[!htbp]
 \begin{center}
  \includegraphics[width=\hsize]{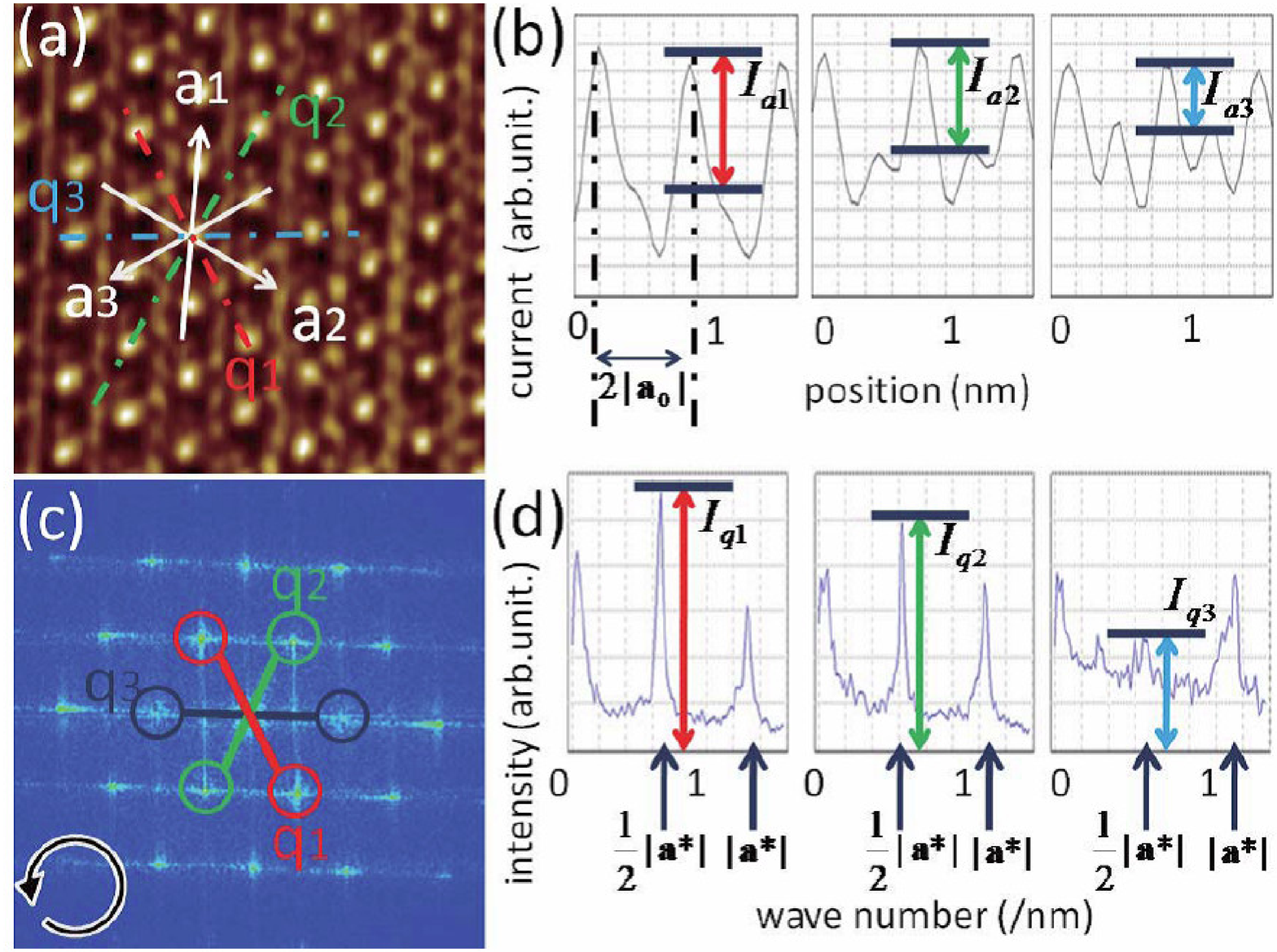} 
  \caption{(a) 3.5 nm $\times$ 3.5 nm STM images and (b) line profiles along unit vector ${\bf a_1}$, ${\bf a_2}$, and ${\bf a_3}$. $I_{ai}$ is the amplitude of the line profiles along ${\bf a_i}$. Sample A was cleaved in 10$^{-8}$ Torr at 77 K and measured in situ at a sample bias voltage of V$_s$ =150 mV and a initial tunnelling current of I$_t$=0.3 nA at T=84 K. (c) FFT image of sample A and (d) line profiles along ${\bf q_1}$, ${\bf q_2}$, ${\bf q_3}$ wave vector. $I_{qi}$ is the intensity of the line profiles along ${\bf q_i}$. FFT was performed over the entire field of view (25 nm $\times$ 25 nm) of STM image.}
  \label{STM}
 \end{center}
\end{figure}

Figure \ref{STM}(a) and (b) show STM images and line profiles along unit vectors, respectively. Figure \ref{STM} shows the low symmetry structure of the electron distribution in TiSe$_2$ with atomic resolution in sample A. In Fig. \ref{STM}(a), the bright area is the point at which many electrons are tunnelling. We observed a triangular lattice of Se atoms with a lattice constant {\it a$_0$}. In addition, we saw a 2{\it a$_0$} wavelength CDW along the unit vectors. To compare the amplitude of the CDW intensity for each nesting direction, we show the line profiles of the tunnelling currents along the unit vectors \mbox{\boldmath $a_1$}, \mbox{\boldmath $a_2$}, and \mbox{\boldmath $a_3$} (Fig. \ref{STM}(b)). In Fig. \ref{STM}(b), there is obvious difference in CDW amplitude among each directions of the unit vector. The CDW amplitude $I_{ai}$ was specified for each magnitude and labeled in order starting with largest amplitude. In sample A, the CDW amplitude $I_{ai}$ become $I_{a1}:I_{a2}:I_{a3} = 1:0.7:0.5$. As a result, this STM image does not have the 3-fold symmetry owing to a triangular lattice structure. However, when we obtain line profiles along the unit vector, the profiles reflect the contribution of two CDWs. To clarify the contribution of the CDW, we performed Fourier Transformation (FT) of the STM image.

Figure \ref{STM}(c) and (d) shows an FT image of sample A and line profiles along the nesting wave vectors, respectively. Figure \ref{STM}(c) shows the anisotropic states of CDW more significantly. The bright spots correspond to the intensity peak of each wave vector. The outer intensity peaks are the Bragg peaks of the selenium lattice whose wavelength is $a_0$. The inner peaks correspond to a CDW satellite which wavelength is $2a_0$. Note that the tunnelling current image depends on the bias voltage, which makes difference between FT of the STM image from diffraction pattern such as x-ray or electron. In Fig. \ref{STM}(d), there is no difference in the intensity of Bragg peaks which correspond to the each \mbox{\boldmath $a^*$} vector. In contrast, we can see an obvious difference between the three sets of CDW intensity peaks. Then, we denote the wave vectors \mbox{\boldmath $q_1$}, \mbox{\boldmath $q_2$}, and \mbox{\boldmath $q_3$} in order starting with a highest intensity too (Fig. \ref{STM}(d)). In sample A, the CDW intensity $I_{qi}$ becomes $I_{q1}:I_{q2}:I_{q3} = 1:0.9:0.5$. In terms of the rotation in which the indexed number increase (inset in Fig.  \ref{STM}(c)), we name the phase \textit{anticlockwise} phase. An analysis of the Fourier transformation is very important for revealing the anisotropy of CDWs in 2D systems.

\begin{figure}[!htbp]
 \begin{center}
  \includegraphics[width=\hsize]{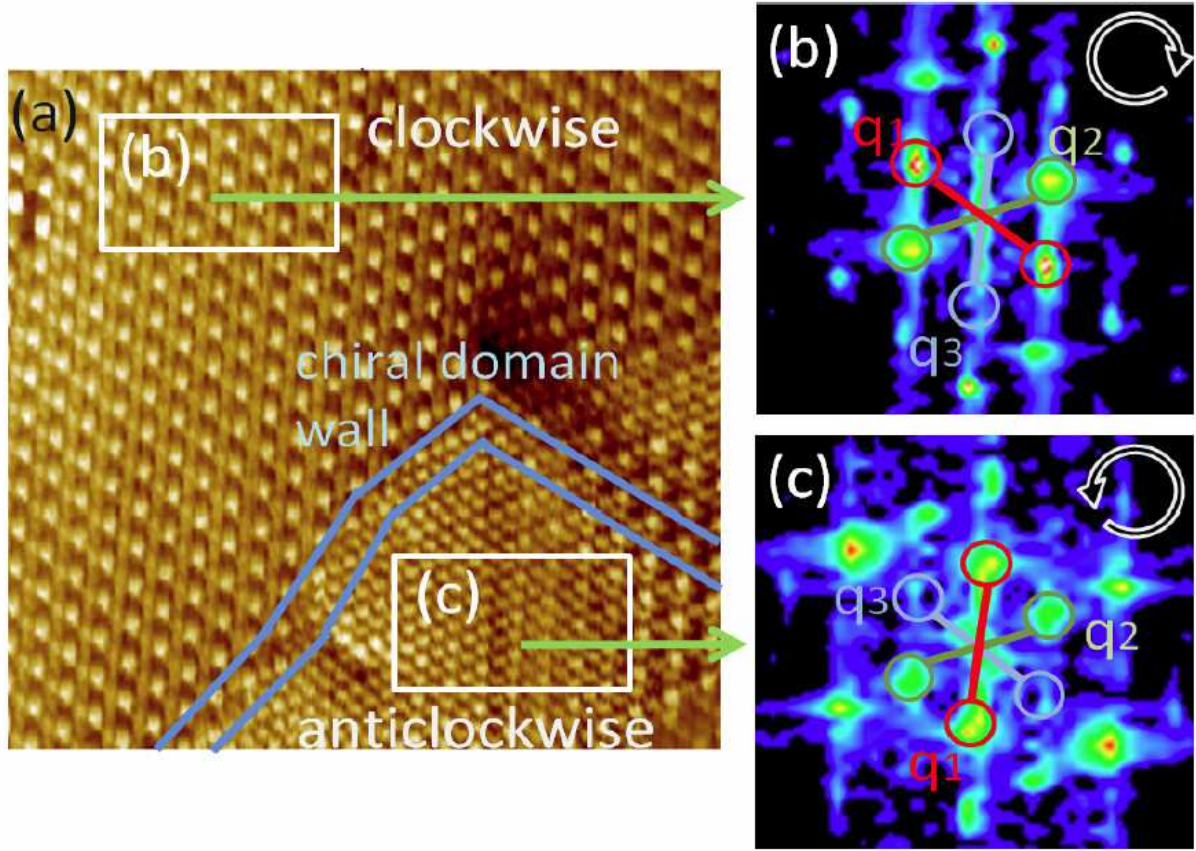} 
  \caption{(a) 12 nm $\times$ 12 nm STM image and (b) (c) FT of two areas each enclosed with a square in sample B. Sample B was measured at V$_s$= -350 mV and I$_t$= 0.4 nA at T= 6.3 K. The blue line is the domain wall between the \textit{clockwise} phase and the \textit{anticlockwise} phase.}
  \label{FFT}
 \end{center}
\end{figure}

Figure \ref{FFT}(a) shows 12 nm $\times$ 12 nm STM images in sample B and figure \ref{FFT}  (b) and (c) show the FFT of two areas each enclosed with a square . In Fig. \ref{FFT}(a), the \textit{clockwise} and \textit{anticlockwise} phases were observed simultaneously. In terms of the rotation in which the indexed number increase as mentioned above, Fig.  \ref{FFT}(b) is \textit{clockwise} and Fig.  \ref{FFT}(c) is \textit{anticlockwise} phase, that is, Fig. \ref{FFT}(b) is a mirror image of Fig. \ref{FFT}(c). Therefore, Figure \ref{FFT}(b) cannot be superimposed on Fig. \ref{FFT}(c) solely with rotational transformation. Thus, the results show the parity symmetry breaking. Although Slough {\it et al}. \cite{STM1} and Coleman {\it et al}. \cite{STM2} observed anisotropy in the conductive plane in a previous STM measurement, the mirror image relation was not demonstrated within their analysis by contour plotting. In this report, we discovered two phases with different chirality. The blue line in Fig. 2(a) shows the boundary between the \textit{clockwise} and \textit{anticlockwise} phases. There is no dislocation around the line in the lattice structure so this is not a crystallographic defect in the observed layer. Thus, the boundary between the two phases is a chiral domain wall. 

From the results of STM measurement to here, we have shown the chiral phases of the CDW in TiSe$_2$. In general, the tunnelling current can obtain information about one or two atomic layers. We employed optical ellipsometry to investigate more macroscopic property of the CDW system.

\begin{figure}[htbp]

    \includegraphics[width=\hsize]{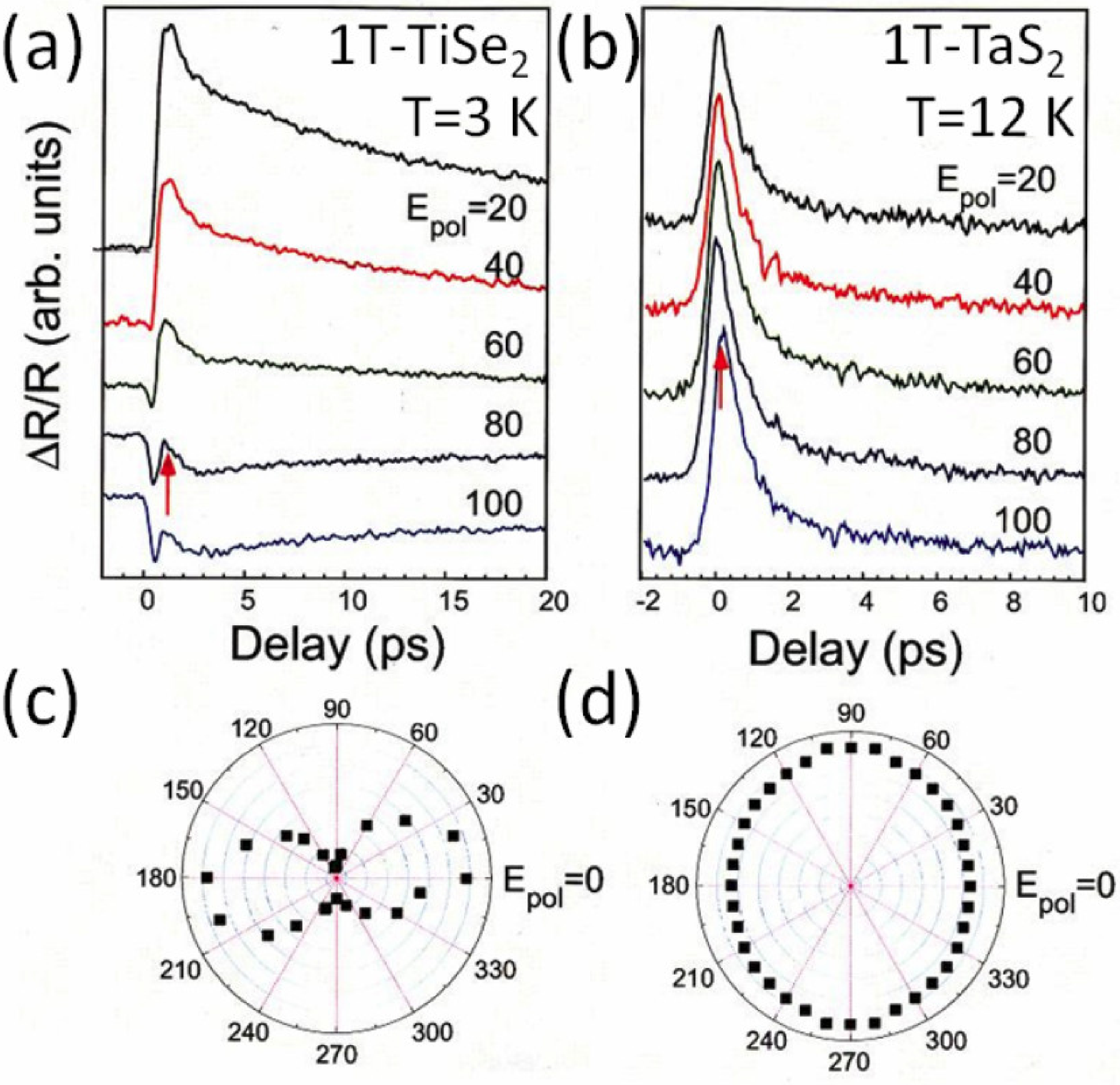} 

  \caption{Polarization dependence of transient reflectivity variation in 1T-TiSe$_2$ measured at 3 K (a) and 1T-TaS$_2$ measured at 12 K (b). The height of each peaks are plotted in the figure (c) and (d) as function of polarization angle $E_{pol}$. The pump pulse wavelength is 1160 nm and the power is 150 $\mu$W. The probe pulse is 800 nm, 80 $\mu$W.
}
  \label{ellipso}
\end{figure}

Figure \ref{ellipso} (a) and (b) show the polarization dependence of the transient reflectivity variation of 1T-TiSe$_2$ measured at 3 K and 1T-TaS$_2$ measured at 12 K, respectively. Figure \ref{ellipso} shows the clear 2-fold symmetry in TiSe$_2$. The transient variation in optical reflectance $\Delta R(t)$ was measured by changing the polarisation angle $E_{pol}$ of the incident probe laser pulse. In general, the peak intensity of transient signals corresponds to the number of electrons excited over the CDW gap \cite{opt,PPmethod}. But in Fig.  \ref{ellipso}(a), the peak intensity behaved anisotropically. Surprisingly, when we plot the peak intensity as function of polarisation angle $E_{pol}$ (Fig.  \ref{ellipso}(c)), we observed 2-fold symmetry although 1T-TiSe$_2$ has a trigonal lattice structure. In contrast, this clear 2-fold property is not seen with 1T-TaS$_2$, which is the most common 2D CDW material with the same lattice structure (Fig.  \ref{ellipso}(b)(d)). Most two-dimensional CDWs do not exhibit an obvious 2-fold symmetry because three CDWs generate a triangular lattice that has 3-fold and 2-fold symmetry simultaneously. Thus, at least within the light penetration length, a special 2-fold structure was achieved in the CDW system. The macroscopic 2-fold property is consistent with the microscopic structure observed with STM.

\begin{figure}[!htbp]
 \begin{center}

  \includegraphics[width=\hsize]{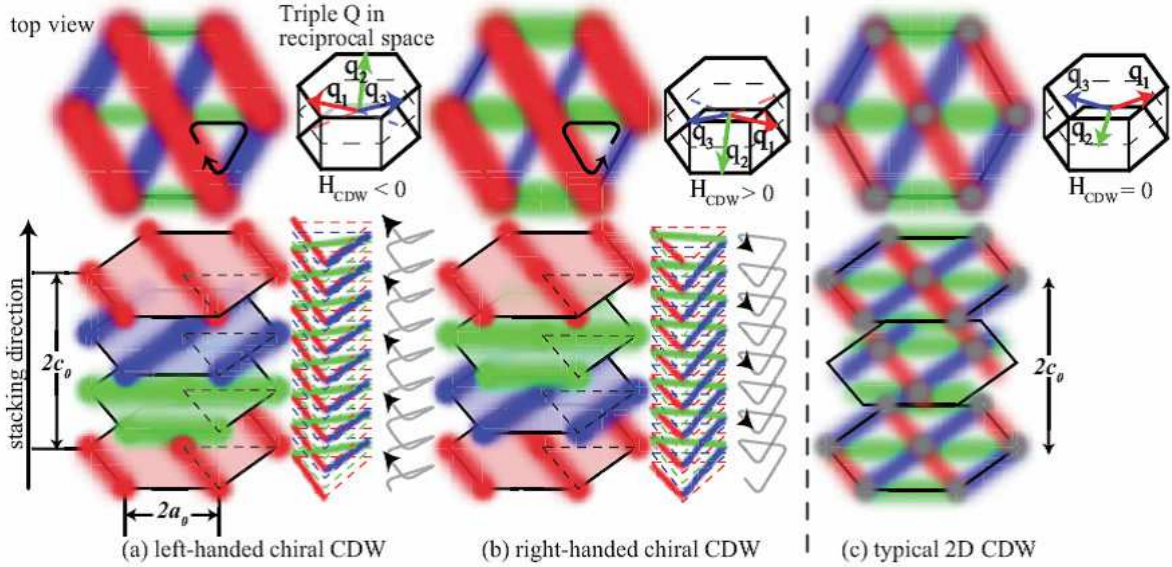} 
  \caption{Schematic representation of (a) left-handed chiral CDWs and (b) right-handed chiral CDWs in a TiSe$_2$ CDW unit cell in real space and (c) typical 2D CDW. Charge concentration is indicated by the deeply coloured part. The colours correspond to the coloured nesting vector in the inset. In a CDW unit cell, the density peaks of three CDW are shifted at intervals of $2c_0/3$. If we look at one layer, the different intensities of three CDWs form a low symmetry structure. There are two phases caused by difference in stacking direction (red-blue-green or red-green-blue). As with cholesteric liquid crystals, the stacking results in the helicity of CDW systems. There is a mirror image relation between the two configurations in (a) and (b). As well as the stacking of CDWs layers in real space, CDW helicity defined in reciprocal space, $H_{CDW} \equiv {\bf q_1} \cdot ({\bf q_2 }\times {\bf q_3})$ reverses by space inversion (${\bf q_i} \to (-{\bf q_i}$)). $H_{CDW}$ is positive in a right-handed chiral CDW, and $\chi _{CDW}$ is negative in a left-handed chiral CDW.}
  \label{model}
 \end{center}
\end{figure}

To understand these results, here let us consider the static charge distribution in real space. We propose a simple model of the CDW stacking. The CDW stacking is not conventional layer stacking with 2D CDW sheets \cite{tanda,tanda2,nakanishi} but CDW component stacking that makes internal structure in a CDW unit cell. Figure \ref{model} shows the configuration of three CDW components. Charge concentration is indicated by the deeply coloured part. Each color corresponds the nesting vector illustrated in the inset of Fig. \ref{model}. In this model, the phase shifting of the density waves along the $c$-axis was permitted, so there is a nonzero relative phase difference between the three density waves. In a CDW unit cell, there are three density peaks at intervals of $2c_0/3$. Chirality is achieved by twisted stacking such as with cholesteric liquid crystals \cite{Chaikin}. Thus, CDW component stacking results in \textit{helical} stacking and helical axis parallel to the $c$-axis. Note that by putting three density peeks in $2c_0$, there are two ways of CDW stacking, that is, when we permutate the sequence of two colored density waves (blue and green), the helix is inverted (Fig. \ref{model}(a)). A Charge helix along the c-axis was achieved in the CDWs and the helicity difference results from the stacking sequence.

This configuration can explain the low symmetry charge distribution in the STM measurement and the 2-fold symmetry in the optical measurement. In the STM measurement, the tunnelling current can obtain information about one or two atomic layers. In one atomic layer, the intensity of three CDWs differs due to the distance between the observed surface and each CDW peaks. A CDW whose peak was closer the surface was more clearly observed ($I{q_1}$) than a CDW that was further away ($I{q_3}$). In terms of the intensity difference between $I{q_1}$ and $I{q_3}$, the FFT image obtained with STM (Fig. \ref{FFT}) reveals the direction of the helical stacking. Along the positive direction of the c-axis, \textit{clockwise} is the ``left-handed" and \textit{anticlockwise} is the ``right-handed" state. The phase shift in one layer as in a conventional 2D CDW cannot explain only the 2-fold symmetry structure. Therefore, the optical results provide evidence for an inter-layer phase shift.

To consider the origin of the \textit{helical} stacking of CDW, we take account of the electronic structure of 1T-TiSe$_2$ in reciprocal space. In contrast to a typical 2D CDW system such as 1T-TaS$_2$, the nesting vectors of TiSe$_2$ originally have a $c$-axis component \cite{band}. In conventional 2D CDW materials such as 1T-TaS$_2$, at first the in-plane triple-\mbox{\boldmath $q$} nesting vectors forms a hexagonal superlattice structure in a S-Ta-S layer. Then, by the Coulomb interaction, the S-Ta-S layers stack with the next CDW layer so as no two density peaks have the same position (Fig. \ref{model} (c)). The phase shift in the neighboring layer results from inter-layer coupling. On the other hand, in 1T-TiSe$_2$, the 3D nesting vectors, that originally has a $c$-axis component, induce a charge modulation directly along the nesting vectors. Accordingly, in 1T-TiSe$_2$ there is no conventional \textit{layer} stacking as in 1T-TaS$_2$. Moreover, relative phase shifting along the $c$-axis between \mbox{\boldmath $q_1$}, \mbox{\boldmath $q_2$}, and \mbox{\boldmath $q_3$} can reduce the Coulomb interaction. As a result, the configuration achieves a \textit{helical} structure (Fig.  \ref{model}).

From above discussion, we propose generalized CDW chirality in triple-\mbox{\boldmath $q$} in reciprocal space.
\begin{equation} 
H_{CDW} \equiv \mbox{\boldmath $q_1$} \cdot (\mbox{\boldmath $q_2$}\times \mbox{\boldmath $q_3$})
\end{equation} 
, where $\mbox{\boldmath $q_1$}$, $\mbox{\boldmath $q_2$}$, and $\mbox{\boldmath $q_3$}$ are the nesting vectors. If the $H_{CDW}$ is zero (three nesting vectors in one atomic plane), a triple-\mbox{\boldmath $q$} CDW exists simultaneously in one layer in a CDW unit cell. But if the $H_{CDW}$ is nonzero (inset in Fig. \ref{model}), there is a degree of freedom owing to charge helicity, that is, $H_{CDW}<0$ represents a left-handed chiral CDW and $H_{CDW}>0$ represents a right-handed chiral CDW. The space inversion of three nesting axes inverts the sign of $H_{CDW}$ (right-handed to left-handed) (Fig. \ref{model}). If we view the CDW system macroscopically, there is chirality. Therefore, we predict that there will be CDW component stacking in a CDW system where $H_{CDW}\neq 0$ and therefore the CDWs themselves will have chirality. 

In conclusion, we discovered chiral CDWs in TiSe$_2$ by using STM and optical measurement. We predict that a chiral electron structure is achieved in 3D nesting CDWs. CDWs lose their 3-fold symmetry with helical stacking and the CDW system achieves chirality. There is a new degree of freedom in a CDW system.

Originally, the phase transition mechanism in TiSe$_2$ had remained a mystery. Several models such as an excitonic insulator \cite{excitonic} or the band Jahn-Terrer effect\cite{Jahn} attempted to explain the unique transition in terms of electronic structure in {\it k}-space, and experimental results have not validated these models completely. Our discovery may help to solve this problem. That is, the anomalously broad peak with a width of 100 K in the resistivity in the phase transition \cite{DiSalvo} may be a consequence of the frustration of left-handed and right-handed phases.

\begin{acknowledgments}
We thank K. Yamaya and T. Toshima for providing advices on sample preparation. We also thank K. Inagaki, T. Matsuura and H. Nobukane for measuring the electrical property. And we thank N. Hatakenaka for helpful comments. This work was supported by the 21COE program on "Topological Science and Technology" from Ministry of Education, Culture, Science and Technology of Japan. 
\end{acknowledgments}


\newpage

\end{document}